\documentstyle[sprocl,epsf,fleqn]{article}
\def\hbar{{\mathchar'26\mkern-7muh}}

\renewcommand{\vec}[1]{\mbox{\boldmath$#1$}}
\newcommand{\Dp}[2]{\frac{\partial #1}{\partial #2}}

\newcounter{dafigcounter}
\def\thedafigcounter{\arabic{dafigcounter}}
\newdimen \figleftportion
\newdimen \figrightportion
\newcommand{\partfig}[4]{\refstepcounter{dafigcounter}
\figrightportion=\textwidth
\figleftportion=#3
\advance \figleftportion by \epsfxsize
\advance \figrightportion by -\figleftportion
\advance \figrightportion by -4mm
\noindent
\begin{minipage}[b]{#3}
\mbox{\epsffile{#1.eps}}
\end{minipage}\hfill
\begin{minipage}[b]{\figrightportion}
\label{#1}\footnotesize Figure \thedafigcounter :\ #2
\vspace*{#4}
\end{minipage}}

\newcommand{\widefig}[2]{\refstepcounter{dafigcounter}
\noindent
\begin{minipage}{\textwidth}   
\begin{center}
\mbox{\epsffile{#1.eps}}
\end{center}
\label{#1}\footnotesize Figure \thedafigcounter :\ #2
\end{minipage}\vskip 5mm}

\def\be{\begin{equation}}
\def\ee{\end{equation}}
\def\bea{\begin{eqnarray}}
\def\eea{\end{eqnarray}}

\begin{document}
\title{MICROSCOPIC CALCULATION OF EXPANDING NUCLEAR MATTER}
\author{JENS KONOPKA, HORST ST\"OCKER, AND WALTER GREINER}
\address{Institut f\"ur Theoretische Physik\\
         Johann Wolfgang Goethe-Universit\"at\\
         Postfach 11 19 32\\
         D--60054 Frankfurt am Main, Germany\\
         e-mail: konopka@th.physik.uni-frankfurt.de}
\maketitle\abstracts{Quantum Molecular Dynamics (QMD) calculations 
are used to study the expansion phase in central collisions between 
heavy nuclei. The final state of such a reaction 
can be understood as the result of a
entropy conserving expansion starting from a compact source. The
properties of this hypothetic source, however, are in conflict
with the assumptions used in fireball models. Moreover, this
hypothetical source is not formed in the dynamical evolution of the
system.}

\section{Introduction}
The question how matter behaves under extreme conditions has initiated
a large number of theoretical and experimental investigations~\cite{NASI93}.
In particular central collisions between two heavy nuclei are known to lead 
to a configuration where the density exceeds that of ordinary nuclei and
the momentum distribution does not correspond to that of a Fermi gas at zero
temperature. One speaks about the creation of hot and dense nuclear matter.
There is quite some evidence from hydrodynamical calculations~\cite{Sch93} 
as well as
from the quantum statistical analysis of the fragmenting source in central
collisions~\cite{Kuhn93}
that the hot and dense systems expands and cools along isentropes
in the temperature-density plane. Moreover,
so called fireball models~\cite{Gos78}, 
which assume an (isentropic) expansion from an 
equilibrated source have been applied with some success to describe data.

On the other hand it is still an open question, to which extent the 
system equilibrates before expanding
and hence allows for the application of thermodynamical 
concepts. In this contribution we address the question, to which extend the
final state of a heavy ion reaction can be interpreted as the result of an
isentropic expansion and which conclusions about the high density stage can
be drawn from the knowledge of the final momentum distribution only.

First the basic features of the Quantum Molecular Dynamics model with
Pauli-potential are briefly reported. The source of irreversibility in
microscopic transport models is discussed with this particular example. A
novel method of reverting an isentropic expansion is introduced. Implications
on the applicability of equilibrium models are critically examined.

\section{Quantum Molecular Dynamics with Pauli-Potential}
Quantum Molecular Dynamics (QMD)~\cite{Aic88,Pei89,Pei92}
is a semi-classical model which calculates
the trajectory of a heavy ion collision in the entire many-body phase-space.
It simulates the many-body dynamics due to the real and imaginary part of
the optical potential by merging two transport theoretical approaches:
The real part of the potential is treated via a phenomenological
nucleon-nucleon interaction, whereas the effect of the imaginary part can
be translated into a collision term of the Boltzmann type~\cite{Aic91}.
QMD therefore
contains a classical molecular dynamics section and a collision term, which
performs a Monte-Carlo integration of the local collision kernel in a 
similar manner as in intranuclear cascade models~\cite{Yar79}.

Special attention has been given to the fermionic character of the 
nucleons in the molecular dynamics part as well.
For an antisymmetrized state of two gaussian wavepackets
the expectation value of the kinetic energy operator reads
\be
\label{ekinasy}
E_{\rm kin} = \frac{p^2}{2\mu} + \frac{3\alpha \hbar^2}{4\mu}
+ \frac{\alpha \hbar^2}{2\mu}
\frac{\alpha r^2 + \frac{p^2}{\alpha (\hbar c)^2}}
{\exp\left\{\alpha r^2 + \frac{p^2}{\alpha (\hbar c)^2}\right\}-1} \, ,
\ee
where the second term on the right hand side is due to the zero point 
energy of the wavepackets. Since the width parameter $\alpha$ is 
time-independent and the corresponding energy cannot be transformed into 
other forms of energy this constant term is neglected. The third term can 
be interpreted as a coordinate- and momentum-dependent potential between
the two gaussians~\cite{Wilets77}. 
However, in our case we took another functional form of 
the Pauli-potential~\cite{Dorso87}
\be
V_{\rm Pauli} = V_0^{\rm Pauli}\left(\frac{\hbar}{p_0q_0}\right)^3
            \exp\left\{-\frac{(\vec{x}_j-\vec{x}_k)^2}{2q_0^2}
                       -\frac{(\vec{p}_j-\vec{p}_k)^2}{2p_0^2}\right\}
            \delta_{\tau_j\tau_k}\delta_{\sigma_j\sigma_k} \,,
\ee
whose parameters have been adjusted to the temperature- and 
density-depen\-dence
of the kinetic energies of a free Fermi-gas~\cite{Pei92}. The delta-functions
indicate that this potential acts only between particles with identical
spin and isospin projection.

Taking into account Fermi momenta in such a manner allows for a
selfconsistent determination of nuclear ground-states by searching
for that configuration, which binds $A-Z$ neutrons and $Z$ protons and 
minimizes the total energy. Necessary conditions for this minimum are
\be
\label{EminBed}
\Dp{H^A}{\vec{p}_j} = \dot{\vec{r}_j} = {\bf 0}
\quad {\rm and} \quad
\Dp{H^A}{\vec{r}_j} = -\dot{\vec{p}_j} = {\bf 0} \,,
\qquad j=1,\ldots, A \,.
\ee
From the last equation we can conclude that these model nuclei are absolutely
stable, because none of nucleons moves nore there is force acting on any 
of the constituents.

In the following, we refer to QMD with the collision term switched off as 
molecular dynamics (MD). However, one has to keep in mind that the nucleons
in the model 
do not behave as classical particles. The usage of a momentum-dependent 
potential implies that velocity and momentum of the particles are not
proportional to each other. 

\section{Time-Reversibility of Molecular Dynamics}
In order to prove the reversibility of the molecular dynamics section in
the QMD model, we performed a molecular dynamics calculation of the 
system Au + Au at an incident beam energy of 150 MeV/nucleon and an impact
parameter of 3 fm.
This is a typical heavy ion reaction, which has been studied 
experimentally with the Plastic-Ball more than a decade ago~\cite{Dos85}
and has received novel attention because of the more recent investigations
with the FOPI-detector at GSI~\cite{Ala92}.

The system is propagated on its molecular dynamics trajectory for a
typical collision time of 400 fm/c. Then the momenta of all particles are
mirrored and the system is propagated back with a negative time-step
for another 400 fm/c. Stages during the back-progation are labelled 
with a prime, e.g.\ the instant 160' fm/c corresponds to a
propagation for 400 fm/c with a positive time step width and 
an backpropagation for additional 240 fm/c with
a negative time step width. In case of perfect reversibility the 
particles' positions in
configuration space at that instant should be identical with the situation
after 160 fm/c.
Fig.\ \ref{crismd-md} shows typical snap shots of coordinate-
and momentum-space projections during the molecular dynamics 
simulation. Starting from two initially well separated gold nuclei 
(bottom row, 0 fm/c), 
in the course of the reaction the system partially disintegrates 
into light fragments and a few very heavy clusters, although no hard 
scatterings are involved (middle row, 400 fm/c). 
This kind of fragmentation is exclusively caused 
by the internucleon potentials. After the back propagation is completed
(top row, 0' fm/c) projectile and target have formed again, which illustrates
the time reversibility of MD.

To be more quantitative, the spectrum of distances between the positions
in the initial state and those after the full propagation there and back
have been calculated. 
The mean displacement is less than 1 fm in configuration- and 50 MeV/c
in momentum-space. This is mainly due to too a coarse discretization when
the high densities are reached. 

The results presented have been obtained with default 
model parameters, which are employed in the simulations too. 
Using higher order integration routines and/or
smaller time-steps could improve the agreement even more. However, in view of
the fact that the dynamics at short relative distances will be governed by 
hard collisions, the results of QMD calculations are not affected by this
discretisation error.

\section{Reverting an Isentropic Expansion}
The time-reversal symmetry of the MD calculations shows that system follows 
an isentropic path. It is tempting to use the concept of the back propagation 
in order to time-revert the expansion of a simulated heavy ion collision, 
which is assumed to conserve the entropy
\cite{}. A typical evolution of a single
event is shown in Fig.\ \ref{crisqmd-md}. In contrast to
Fig.\ \ref{crismd-md} the Au + Au system is propagated 
on its QMD trajectories, including the hard scatterings. Hence entropy may
be produced during the compression--decompression dynamics. 
Again, the calculation is carried out for 400 fm/c and the back-propagation
is carried out using MD only.

With collisions included (QMD-MD), 
the time-evolution drastically differs from the
MD-MD calculation. The system fragments more violently, 
the spectators bounce off the
hot and dense participant matter and the typical flow 
ellipsoid develops in momentum space. In contrast to the MD-MD calculation
shown in Fig.\ \ref{crismd-md} the 
QMD-MD simulation nolonger leads back to two well separated nuclei.
All particles seem to stem from a single compact source. In momentum-space,
however, the nonisotropic emission pattern is present, even after the
backpropagation is completed.

The future evolution of a classical
dynamical system in general is strongly dependent on
the distribution of matter in phase-space. In particular the 
initial correlation 
between configuration and momentum space determines the dynamics.

In order to study the physics of the expansion phase in more detail, we 
have calculated the time-evolution of the one-body distribution function in
QMD. Technically this is achieved 
by superposition of many events with the same incident
energy and impact parameter. The result of this procedure is equivalent to 
a testparticle distribution of a VUU/BUU calculation. 
It allows for determination of the local velocity distribution
with arbitrary precision, which is only limited by the number of
events. The first and the second moment of 
the local velocity dsitribution are related to collective motion and the
local temperature respectively.

Fig.\ \ref{wilder_flowtime} compares the distribution of matter
and the local collective motion in the event-plane after the isentropic
backpropagation until the highest density is reached again (left)
with the corresponding stages on the system's true dynamical path (right).

Although the backpropagated state also exhibits a central density of 
twice the saturation density, the correlation between space and momentum
shows a behaviour which is never observed in the ordinary QMD simulation.  
It shows that the final state of a heavy ion reaction can
indeed be understood as the result of an isentropic expansion from a single
compact source. However, the correlation between configuration and 
momentum-space has never really been traversed by the system.

A further analysis of the backpropagated starting point of the isentropic
expansion is displayed in in Fig.\ \ref{pxmd}. The mean
directed transverse momentum as a function of rapidity, which is observed
in the final state of the reaction survives the isentropic backpropagation
almost completely. Only the rapidity distribution widens a bit compared to the
final state. Again it is recognized that the isentropic backpropagation
preserves the collective flow correlations. These correlations are 
just starting to develop in the corresponding temporal stages during
the QMD propagation to the final state, which are shown as lines in 
the figure.

\section{Summary}
In summary, we have shown that the transport mechanism in the QMD model 
can be divided into an entropy conserving part and a stochastic part which
is the underlying microscopic cause of the newly produced entropy in heavy
ion reactions.
The time-reversal symmetry of the classical equations of motion
allows for a reversion of the anticipated isentropic expansion in the 
simulation. 

The hypothetical starting point of the isentropic expansion of nuclear 
matter in heavy ion collisions indicates that all matter stems from
a single compressed source. On the other hand, the position--velocity
correlations observed after the isentropic backpropagation differ
drastically from those present in the corresponding reaction. The 
present analyses suggest that fireball models which assume an isentropic
expansion from a compact source may describe data, if flow correlations 
were taken into account. However, the trajectories do not correspond to
the true dynamical path of the system. This is due the fact that entropy is
produced via abundant two body scatterings not only in the high density phase
but also during the expansion.

\section*{Acknowledgments}
This work has been supported by the Bundesministerium f\"ur 
Bildung und Forschung (BMBF), the Deutsche Forschungsgemeinschaft (DFG), and
the Gesellschaft f\"ur Schwerionenforschung mbH (GSI).

\newpage
\epsfxsize = 9.1cm 
\widefig{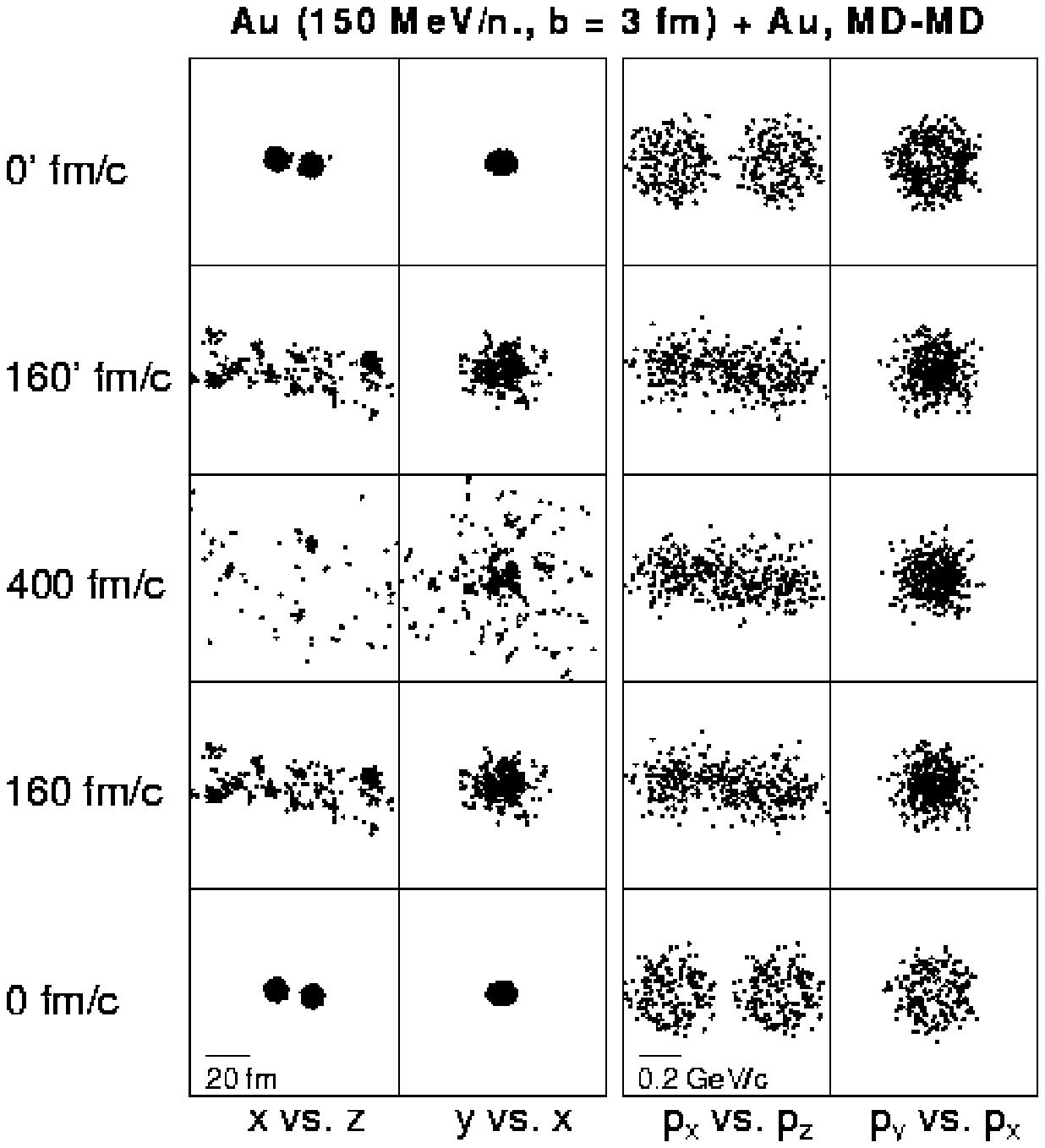}{Position and momenta of nucleons in 
a classical molecular dynamics calculation of the reaction 
Au (150 MeV/nucleon, b = 3fm) + Au. After propagation for 400 fm/c, the 
momenta are mirrored and the system is propagated back for another 
400 fm/c. Projections in configuration space (1st and 2nd column) 
and in momentum space (3rd and 4th column) onto the event plane
(1st and 3rd column) and onto the plane perpendicular to the 
beam axis (2nd and 4th column) are displayed.}

\newpage
\epsfxsize = 9.1cm 
\widefig{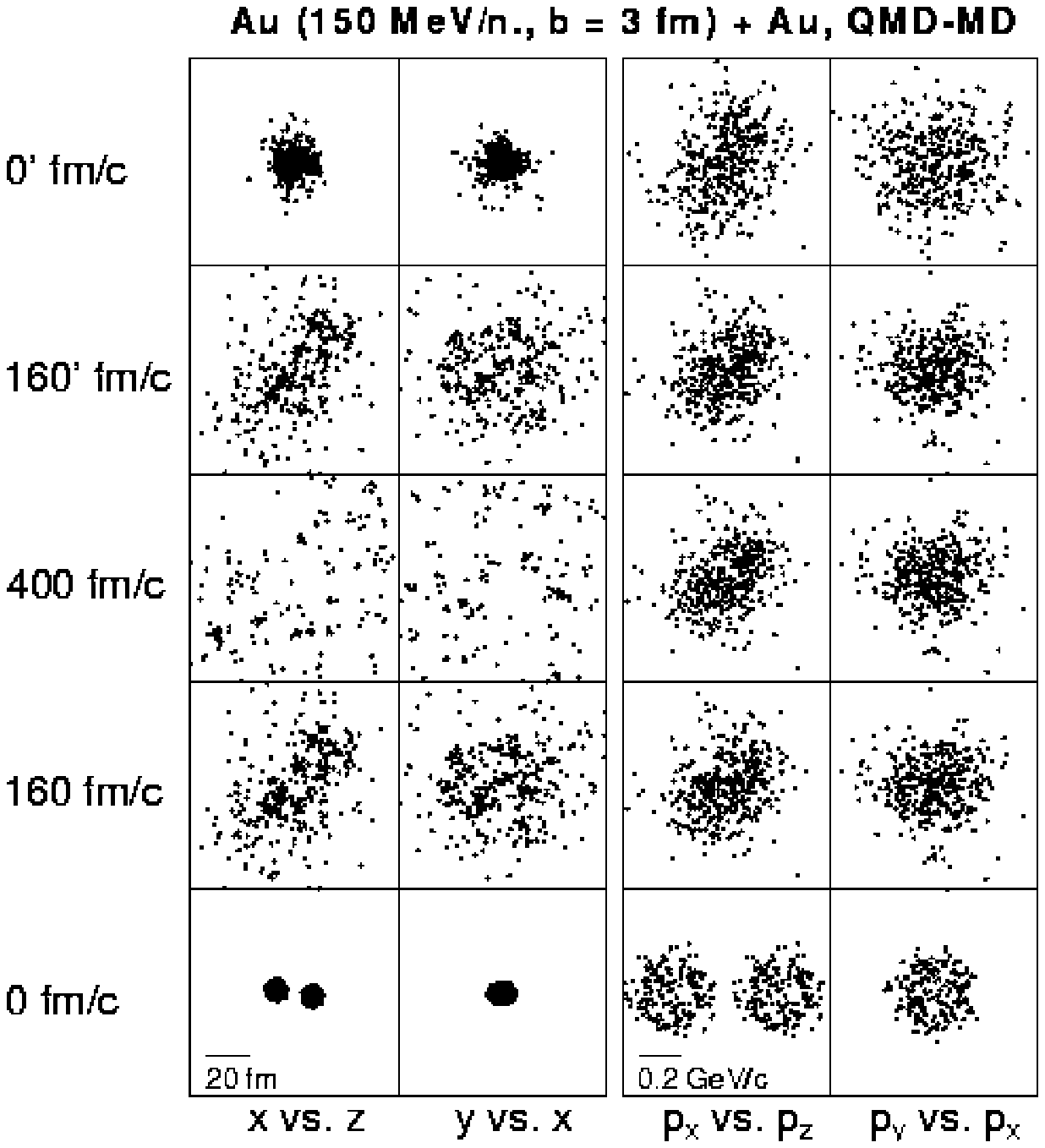}{Same as 
Fig.\ \protect{\ref{crismd-md}} but for a MD back 
propagation after a full QMD propagation.}

\newpage
\epsfxsize = 11.9cm 
\widefig{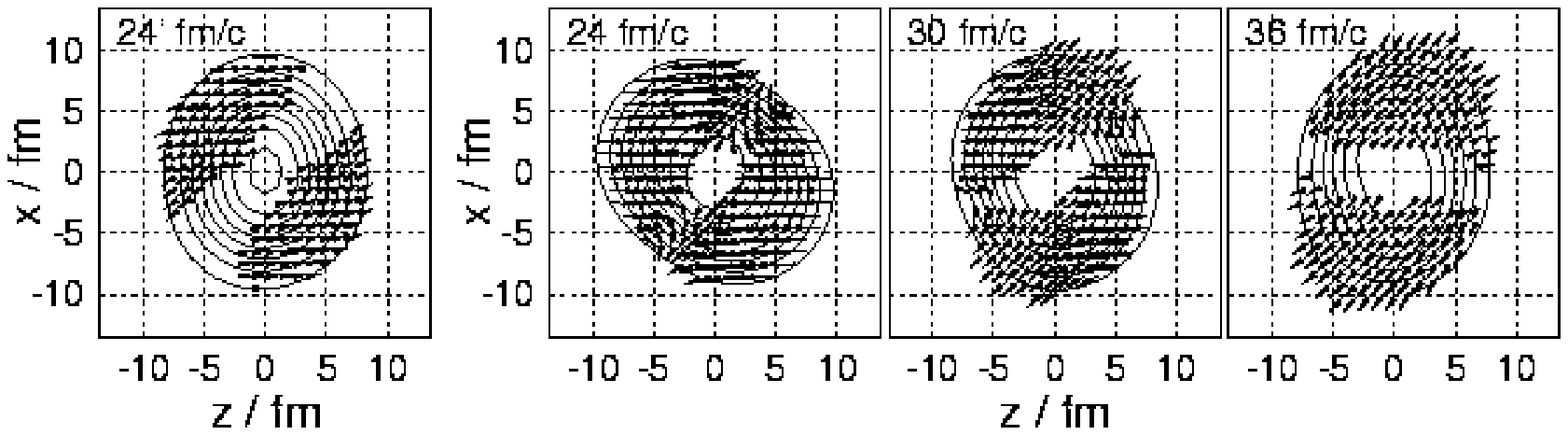}
{The backpropagated hypothetical starting point of an isentropic expansion
(left) is compared to the corresponding reaction stages (right). The results
have been obtained for Au (150 MeV/nucleon, b=3fm) + Au collisions, and show
a cut through the reaction plane. The distance between
two contour lines is 0.2$\varrho_0$. The outermost contour indicates a
density of 0.2$\varrho_0$. The arrows indicate the collective motion. Only
arrows corresponding to velocities larger than 0.05 c are displayed.}

\newpage
\epsfxsize = 5cm 
\widefig{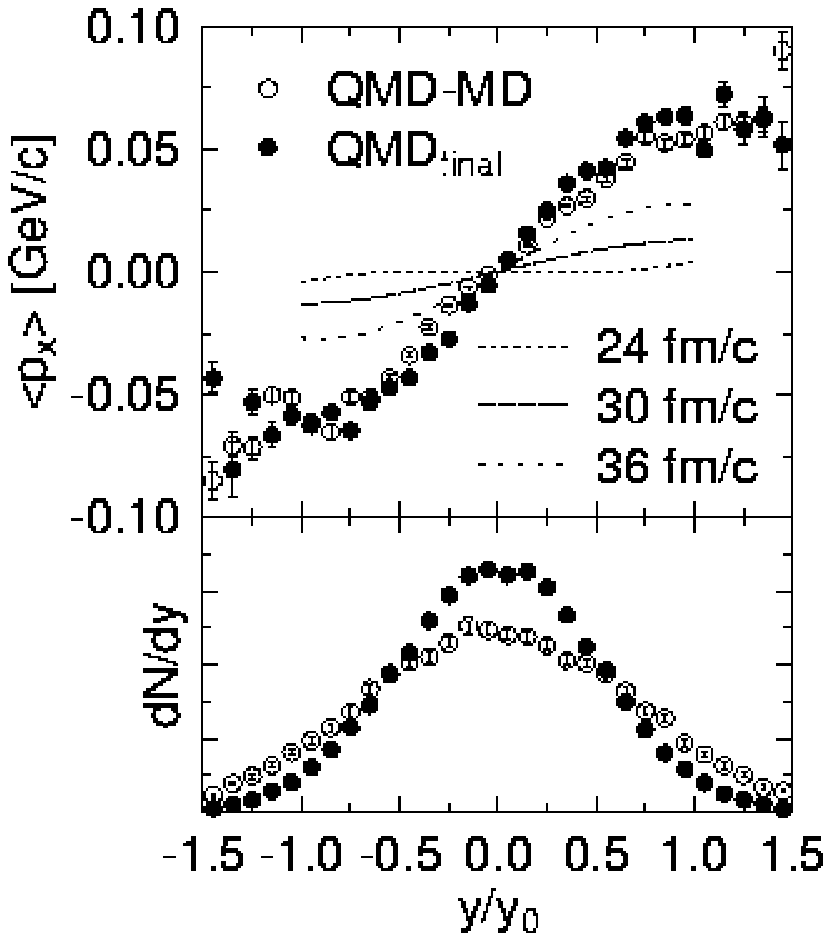}{Flow correlations and rapidity distributions
in the final state (closed circles) are compared to those of the state of 
maximum density in the backpropagation (open circles). The molecular dynamics
trajectories preserve almost completely the flow correlations. Some of the
$\langle$p$_{\rm x}\rangle$(y) 
curves during the QMD propagation are shown as lines.}

\end{document}